\documentclass{ws-procs975x65}

\begin{document}

\title{
TESTING GENERAL RELATIVITY USING GRAVITATIONAL WAVES FROM BINARY NEUTRON STARS: EFFECT OF SPINS}

\author{M.~AGATHOS$^{1*}$, W.~DEL~POZZO$^{1}$,  T.G.F.~LI$^{1}$, C.~VAN~DEN~BROECK$^{1}$, J.~VEITCH$^{1}$, S.~VITALE$^{1,2}$}

\address{${}^{1}$Nikhef--National Institute for Subatomic Physics,\\
 Science Park 105, 1098 XG Amsterdam, The Netherlands\\
${}^{2}$LIGO - Massachusetts Institute of Technology, Cambridge, MA 02139, USA\\
$^*$E-mail: magathos@nikhef.nl}
\begin{abstract}
We present a Bayesian data analysis pipeline for testing GR using gravitational wave signals from coalescing compact binaries, and in particular binary neutron stars.
In this study, we investigate its performance when sources with spins are taken into account.

\end{abstract}


\bodymatter

\section{Introduction}

Among the most promising classes of gravitational wave (GW) sources in the advanced detector era is that of coalescing binary systems, consisting of neutron stars (NS) and/or black holes.
Several parallel efforts within the context of the post-Newtonian (PN) formalism~\cite{Blanchet:2002av}, have yielded very accurate modelling of gravitational waveforms coming from the inspiral
stage of the evolution of compact binaries,
increasing the amount of high-order relativistic effects included in the models and thus providing data analysts with an invaluable toolkit for extracting scientific information from the raw data of the GW detectors.

Given a set of GW signals, one of the most important questions is that of their compatibility with the predictions of general relativity (GR)~\cite{Yunes:2009ke}.
Here we present TIGER (Test Infrastructure for GEneral Relativity), a data analysis pipeline that uses Bayesian model selection to assess the agreement of the data with GR or inconsistency thereof.
The groundwork for this analysis is set in Ref.~\refcite{Li:2011cgsh,Li:2011vxsh}, where a proof of principle, as well as performance under a few non-trivial scenarios are demonstrated.

A few possible concerns arise, simply from the fact that any data analysis technique is based on the available model waveform templates on one hand, and noisy data coming from imperfect detectors on the other.
We enumerate the most important ones, before addressing one of them in the next section and defer the investigation of the rest to a forthcoming paper:
\begin{itemlist}
\item effect of errors in the calibration of the detectors
\item mismatch between different waveform approximants
\item effect of spins 
\item impact of NS finite size/tidal effects of unknown magnitude
\item effect of non-gaussian and/or non-stationary noise in the detectors.
\end{itemlist}

\section{Introducing aligned/anti-aligned spins}
In the foundational papers~\cite{Li:2011cgsh,Li:2011vxsh}, the waveform family TaylorF2~\cite{Buonanno:2009zt} was chosen as a fast and accurate frequency domain template (at least for the purpose of modelling the inspiral stage of binary NS systems), with the restriction however of zero spins.
This waveform can be easily extended to accommodate the contribution of (anti-)aligned spins, starting at 1.5PN order in the phase, as demonstrated in Ref.~\refcite{Kidder:1992fr}.

In the current study, we extend the parameter space in our simulated sources to include (anti-)aligned spins and their effects up to 2.5PN order in the phase.
The simulated dimensionless spin magnitudes $\chi$ follow a normal distribution centered at zero, with $\sigma = 0.05$, as a conservative model, motivated by the spins in the NS binaries observed to date~\cite{O'Shaughnessy:2009dk} and their expected evolution up to coalescence time.

The expected consequence is that for a template family that ignores these effects, a signal coming from a binary system with spinning components would appear to violate GR.
This template inadequacy will result in a widening of the \emph{background}: the distribution of odds ratios for simulated GR sources.
The question that remains, is how well the extension of the template family to include (anti-)aligned spin effects can recover sources with (anti-)aligned (or even arbitrary, precessing) spins.

\section{TIGER} 
The main objective of TIGER is, by processing the available data from the detected signal(s), to produce a single number, that would quantify our posterior belief in the data being consistent with GR.
This quantity of interest is what is known as the \emph{odds ratio} between the $\mathcal{H}_{GR}$ and $\mathcal{H}_{modGR}$ hypotheses.

Combining information from multiple detections into a single and most confident statement is one of the virtues of this pipeline.
The \emph{cumulative odds ratio} for a catalogue of $\mathcal{N}$ sources, using $N_T$ test coefficients, is given by~\cite{Li:2011cgsh,Li:2011vxsh}
\begin{equation*}
^{(N_T)}\mathcal{O}^{modGR}_{GR} = \frac{P(\mathcal{H}_{modGR}|{d_1,\cdots,d_\mathcal{N}})}{P(\mathcal{H}_{GR}|{d_1,\cdots,d_\mathcal{N}})} = \frac{\alpha}{2^{N_T} + 1} \sum_{k=1}^{N_T} \sum_{i_1< \cdots <i_k} \prod_{A=1}^{\mathcal{N}}  {}^{(A)}B^{i_1 i_2 \cdots i_k}_{GR} , 
\end{equation*}
where we define the \emph{Bayes} factor $^{(A)}B^{i_1 i_2 \cdots i_k}_{GR}$ as the ratio of evidences between $H_{i_1 i_2 \cdots  i_k}$ and $\mathcal{H}_{GR}$ for source $A$.
Here the multi-index $i_1 \cdots i_k$ represents the model where the PN phase coefficients $\psi_{i_1},\cdots,\psi_{i_k}$ are set as free parameters that deviate from their GR value, while the rest are consistent with GR.
The evidences for the different hypotheses are then estimated, using the implementation of the \emph{Nested Sampling Algorithm} in the \emph{LIGO Algorithm Library}~\cite{Veitch:2009hd}.

\section{Simulations and Results}

For the purpose of this analysis, we use the approximant TaylorF2 up to and including 3.5PN terms in the phase, both for injection and template waveforms.
We simulate both GR-consistent sources (\emph{background}), and GR-violating sources with a $-10\%$ relative shift in the 1.5PN phase coefficient $\psi_3$ (\emph{foreground}).
The parameters of the simulated sources (masses, location, etc.) are randomly distributed in a realistic way along the lines of Ref.~\refcite{Li:2011cgsh}, with the additional spin magnitudes sampled from a Gaussian as explained above.
The simulated GW signals are then injected into Gaussian, stationary noise, according to the noise curves of aLIGO and AdVirgo.

\begin{figure}[t]%
\begin{center}
 \parbox{2.1in}{\epsfig{figure=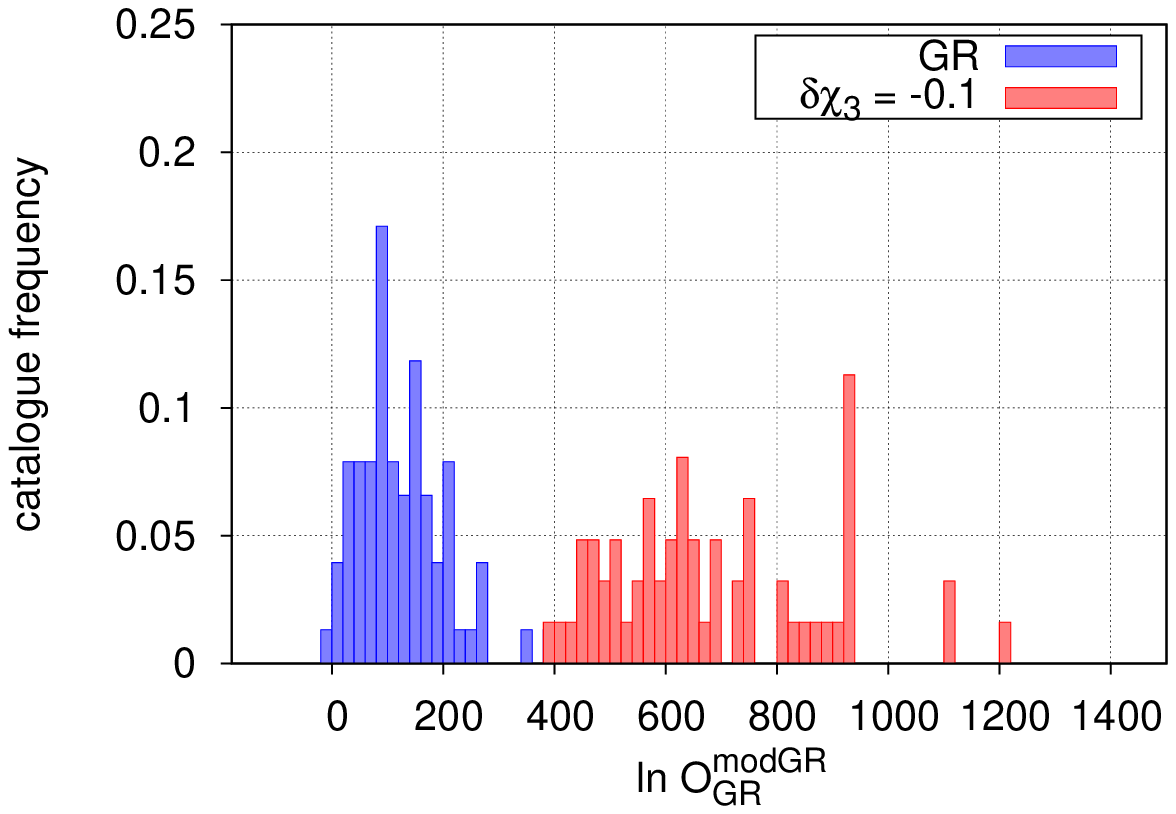,width=2in}
}
 \hspace*{4pt}
 \parbox{2.1in}{\epsfig{figure=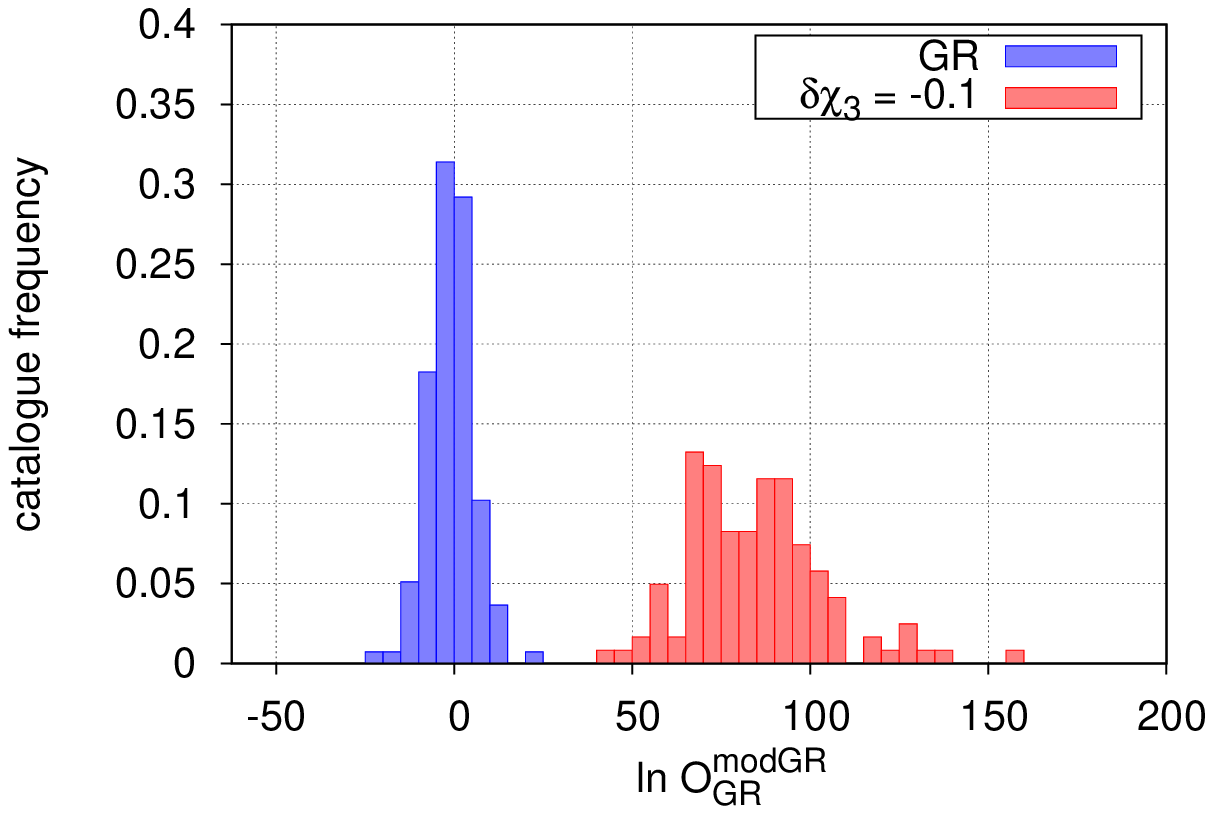,width=2in}
 }
 \caption{Foreground (red) vs background (blue) for catalogues of 15 sources each, using 
 templates with zero spins (left), and templates with aligned/anti-aligned spins (right).\label{results}}
\end{center}
\end{figure}

Subsequently, two different waveform families are used as templates, namely TaylorF2 with and without aligned spins.
The analysis is cut off at $400 \mathrm{Hz}$ in order to suppress any finite size effects that may enter a realistic BNS signal.
The resulting background and foreground distributions of cumulative (log-)odds ratios for catalogues of 15 sources each, are displayed in~\fref{results}. It is clear that the background distribution will not widen significantly, compared to previous results (c.f. Fig.~5 of Ref.~\refcite{Li:2011cgsh}), as long as spin contributions are included in the template waveforms.
Preliminary results point to a similar conclusion for arbitrary, precessing spins.
We also observe a breaking of the degeneracy between spins and GR violations in 1.5PN ($\psi_3$), due to the presence of higher order spin terms in the phase.

\section{Conclusions}

We have investigated the performance of TIGER by analyzing simulated GW signals from BNS systems with aligned spins. 
It has been demonstrated that the spin contributions to the phase can be well distinguished from GR violations, and that a constant $-10\%$ relative deviation in $\psi_3$ is clearly detectable with the first 15 sources.

\bibliographystyle{ws-procs975x65}
\bibliography{bibliography}

\begin{thebibliography}{1}

\bibitem{Blanchet:2002av}
L.~Blanchet, {\em Living Rev. Rel.} {\bf 5}, p.~3 (2002).

\bibitem{Yunes:2009ke}
N.~Yunes and F.~Pretorius, {\em Phys. Rev.} {\bf D80}, p. 122003 (2009).

\bibitem{Li:2011cgsh}
T.~G.~F. Li {\em et~al.}, {\em Phys. Rev.} {\bf D85}, p. 082003 (2012).

\bibitem{Li:2011vxsh}
T.~G.~F. Li {\em et~al.}, {\em J. Phys. Conf. Ser.} {\bf 363}, p. 012028
  (2012).

\bibitem{Buonanno:2009zt}
A.~Buonanno, B.~Iyer, E.~Ochsner, Y.~Pan and B.~Sathyaprakash, {\em Phys. Rev.}
  {\bf D80}, p. 084043 (2009).

\bibitem{Kidder:1992fr}
L.~E. Kidder, C.~M. Will and A.~G. Wiseman, {\em Phys. Rev.} {\bf D47}, 4183
  (1993).

\bibitem{O'Shaughnessy:2009dk}
R.~O'Shaughnessy and C.~Kim, {\em Astrophys.J.} {\bf 715}, 230 (2010).

\bibitem{Veitch:2009hd}
J.~Veitch and A.~Vecchio, {\em Phys. Rev.} {\bf D81}, p. 062003 (2010).

\end{thebibliography}

\end{document}